\documentclass{elsarticle}

\usepackage{subfigure}
\usepackage{graphicx}
\usepackage{epsfig}
\def \neq {\ensuremath{{\rm n}_{eq}/{\rm cm}^2}}
\begin{document}

\title{ Silicon Photo-Multiplier radiation hardness tests with a beam controlled neutron source.}

\author{M. Angelone and M. Pillon}

\address{ENEA, Frascati, Italy}

\author{R. Faccini$^{\dagger,*}$, D. Pinci$^*$}
\address{$^\dagger$ Sapienza Universit\`a di Roma, $^*$ INFN, Sezione di Roma, Italy}


\author{W. Baldini$^*$, R. Calabrese$^{*,\dagger}$, G. Cibinetto$^{*,\dagger}$, \\
A. Cotta Ramusino$^*$, R. Malaguti$^*$, and M. Pozzati$^\dagger$.}

\address{$^\dagger$ Universit\`a degli Studi di Ferrara, $^*$  INFN Ferrara, Italy}

\begin{abstract}
We report radiation hardness tests performed at the Frascati Neutron Generator on 
silicon Photo-Multipliers, semiconductor photon detectors 
built from a square matrix of avalanche photo-diodes on a silicon substrate.  
Several samples from different manufacturers have been irradiated 
integrating up to 7$\times$10$^{10}$ 1-MeV-equivalent neutrons per cm$^2$. 
Detector performances have been recorded during the neutron irradiation and a 
gradual deterioration of their properties was found to happen already after 
an integrated fluence of the order of 10$^8$ 1-MeV-equivalent neutrons per cm$^2$.

\end{abstract}
\maketitle



\section{Introduction}
Silicon Photo-Multiplers ~\cite{bib:irst,bib:hama}  are constituted of a large number of micro-pixels each made of  an APD counter in series with a quenching resistance and they operate in Geiger mode. Their sensitivity to a small number of photo-electrons and their fast response make them candidate light detectors also for the 
extruded scintillators of the Instrumented Flux Return (IFR) of  
the super flavor factory experiment proposed at ``Laboratori Nazionali di Frascati'' (SuperB~\cite{bib:superb}). 
Extremely high luminosities are to be achieved at SuperB at the cost of very high backgrounds, among which neutrons impacting on the detectors. 
While the radiation hardness to photons and charged particles has been studied in detail~\cite{radhard}
the knowledge of the impact of neutron irradiation on Silicon Photo-Multipliers is based on preliminary results on nuclear reactor tests~\cite{reactor}. Due to the impossibility to record data during irradiation and to control the neutron energy and flux, these tests can only assess that the devices are severely damaged after an integrated fluence as high as $10^{11}\neq$, roughly one year of SuperB data-taking without appropriate shielding. No information is available on the maximum fluence that can be absorbed, i.e. on the behavior of the devices for intermediate fluences.
\begin{table}
\begin{center}
\caption{Distance of the devices from the neutron generation point, bias voltage during the irradiation, baseline dark currents and total integrated  and average differential fluences. The conversion to the  
fluence equivalent to 1 MeV neutrons on silicon (n$_{eq}$) is based on Ref.~\cite{bib:std}.}
{\begin{tabular}{lcccccc}
\hline
Device & d   & $V_{bias}$ (V)& $I_{dark}^0$($\mu$A) & tot. int. fluence & avg. diff. fluence\\
       &(mm)&&& (10$^{10}\neq$)  & (10$^{6}\neq/$s)  \\
\hline
SiPM \#4 & 6.0 & -33.0 & 1.2 & 1.25 & 0.9 \\
SiPM \#6 & 4.4  &-33.0& 0.3 & 3.07  &2.1\\
SiPM \#7  &13 & -33.5&  0.3 & 0.18 &0.3\\
SiPM \#8 &3.3 & -33.0& 0.7 & 7.32 &5.0\\
MPPC \#5 &2.5& -70.0& 3.2 & 4.26 &6.7\\
MPPC \#6 &2.5& -70.0& 1.6 & 4.26 &6.7\\
SiPM 2$\times$2 &6.1& -33.0 & 6.1 & 1.25&0.9 \\
\hline
\end{tabular}}
\label{tab:fluences} 
\end{center}
\end{table}
\vspace{-20pt}

\section{Measurement Setup}
\begin{figure}[hbt]
\begin{centering}
\psfig{file= 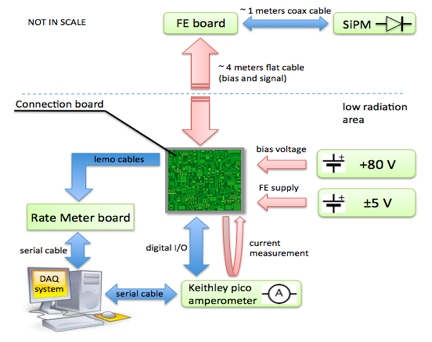, width=10cm}
\caption{Scheme of the supply and readout system of the experiment}
\label{fig:setup}
\end{centering}
\vspace{-5pt}
\end{figure}
This paper reports the results from irradiation tests on a neutron source, the "Frascati Neutron Generator" (FNG~\cite{bib:fng}), that uses a deuteron beam accelerated up to 300 keV impinging on a deuteron target 
to produce a nearly isotropic 2.5 MeV neutron output via the D(d,n)$^3$He fusion 
reaction. The beam current at the target can be regulated 
in order to obtain up to a maximum neutron production rate of 5$\times$10$^8$ neutrons per second on the whole solid angle. During our tests, the neutron yield was monitored online by measuring the rate of recoiling protons measured with a  calibrated liquid scintillator (NE213). Pulse shape discrimination is used to reject gamma-ray events. The temperature of the experimental hall was also monitored and found to be stable between 23 and 25 degree Celsius, stability achieved by means of air conditioning.
The Monte Carlo neutron and photon transport code MCNP-5~\cite{MCNP5} was used to perform a full simulation of the whole experimental hall and to convert the neutron yield into the flux impinging on the detector, including also secondary effects like the neutron scattering  from the bunker concrete walls. An "ad hoc" source routine was also used in the MCNP code to accurately simulate  the source anisotropy arising from the beam-target interaction.

Five devices (four 1$\times$1 mm$^2$ and one 2$\times$2 mm$^2$) produced by the Istituto di Ricerca Scientifica e Tecnologica (IRST~\cite{bib:irst}), named SiPM in the following, and two 1$\times1 \rm{mm}^2$ produced by the Hamamatsu\cite{bib:hama}, named
MPPC in the following, have been irradiated
with neutrons. Their position with respect to the neutron generation point and the corresponding integrated fluence, delivered in six runs of shortly more than one hour each, is shown in 
Tab.\ref{tab:fluences}.

\begin{figure}[hbt]
\begin{centering}
\psfig{file= 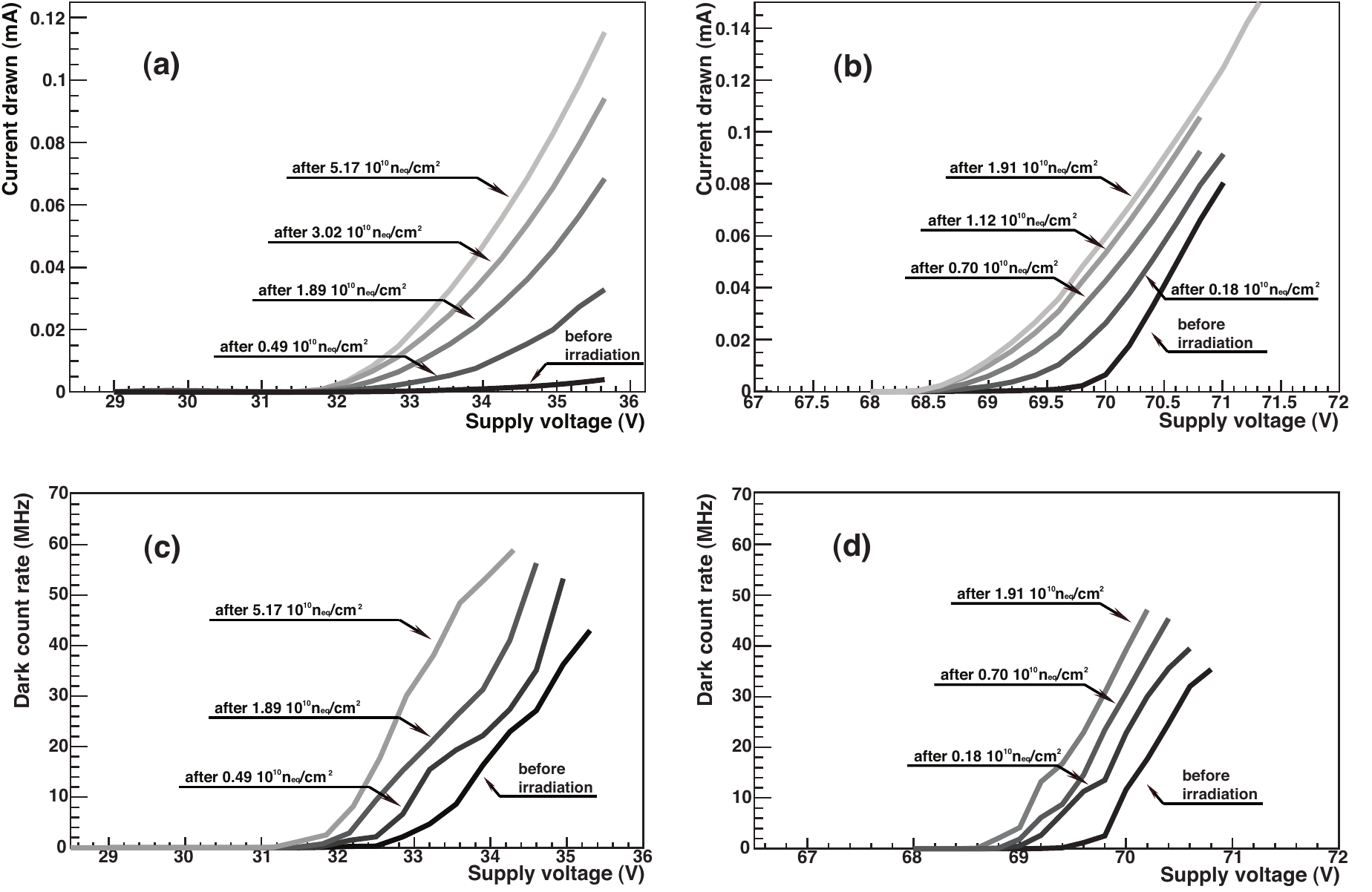, width=10cm}
\caption{Behavior of the I-V characteristics and the rates for a 2.5 photo-electron threshold as a function of the bias for a SiPM (a, c) and an MPPC (b, d) after different integrated fluences.}
\label{fig:lvscan}
\end{centering}
\vspace{-5pt}
\end{figure}

  During the exposure the devices were biassed and their current read in turn by a pico-amperometer driven by a relay as in the scheme in  Fig.~\ref{fig:setup}:  the SIPM were biassed between 3V and 4V above the breakdown bias, while the MPPC were  biassed  at the operation point. At the end of each irradiation run a scan in bias was performed in order to measure the $I-V$ characteristic.
  The bias voltage was distributed by the frontend boards, placed at about one meter from the neutron target, which also provided signal amplification and discrimination for the rate measurement. The bias was provided by amplifying the signal from a LM336/2.5 reference diode with an operational amplifier and under the operational conditions, including temperature, was stable in voltage within 3\%, which translates into a stability within few tens of mV in the bias. The output resistence of the system is 10$k\Omega$ and therefore the variation of the current during the measurements generates a change in bias of at most 100mV.
  
  The data acquisition system allowed to monitor currents, rates, and biases continuously and to change the settings from remote without accessing the experimental hall. Photodetectors temperature, neutron rate and integrated fluence were also recorded for offline analysis.

\begin{figure}[hbt]
\begin{centering}
\psfig{file= 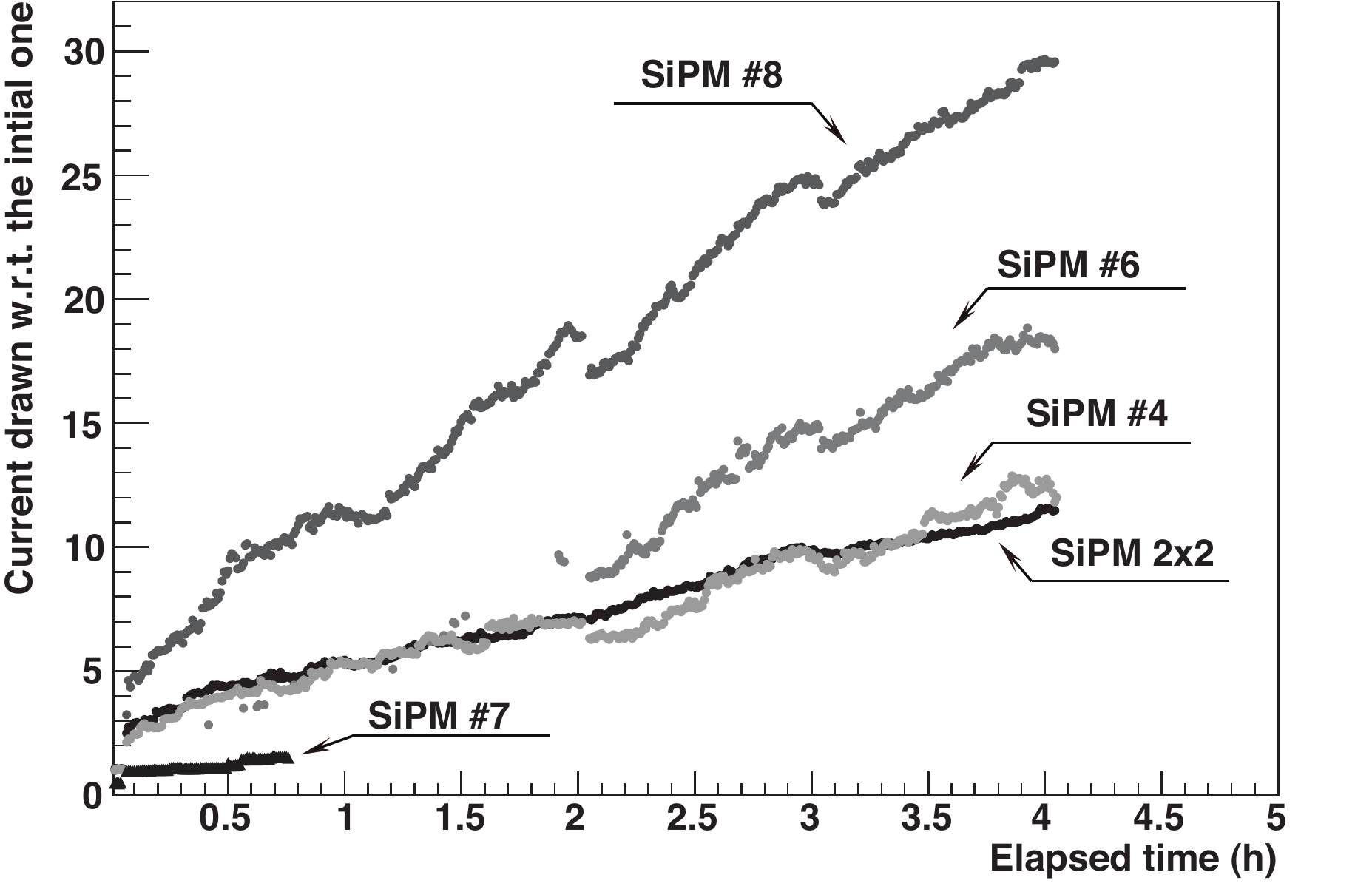,width=10cm}
\caption{ Increase factors of the current drawn by the SiPMs as a
function of the irradiation time.
\label{fig:curr_time}}
\end{centering}
\end{figure}

\section{Currents and rates}

Figure~\ref{fig:lvscan} superimposes the measured characteristics taken after different fluences of irradiation for a SIPM and an MPPC. It shows that the drawn current increases significantly even for relatively small fluence ($5 \times 10^9\neq$). The same effect can be observed in the measured rates by setting a threshold at 2.5 photo-electrons.
\begin{figure}[hbt]
\begin{centering}
\subfigure[]{\psfig{file= 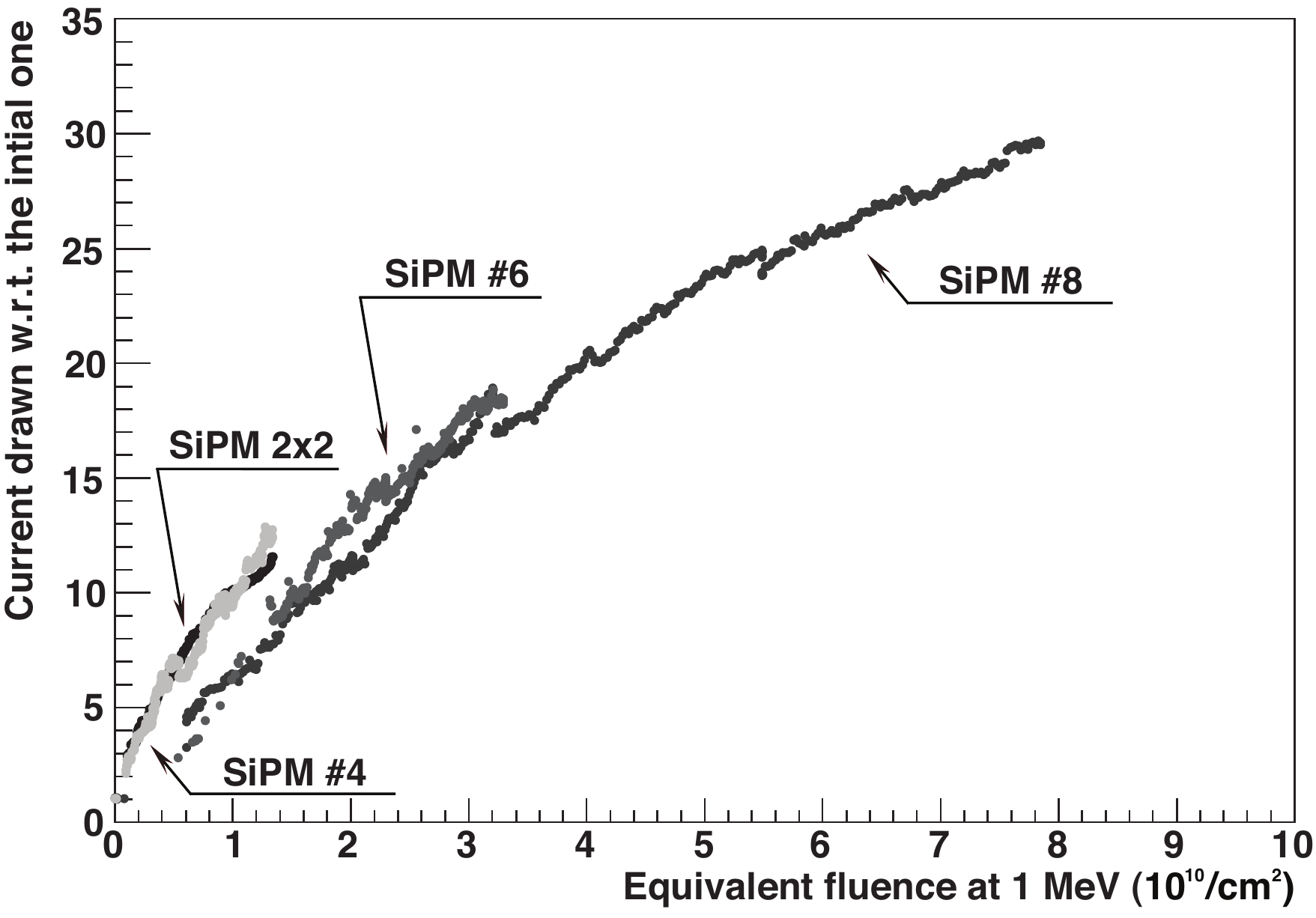,width=8cm}
\label{total}}
\subfigure[]{\psfig{file= 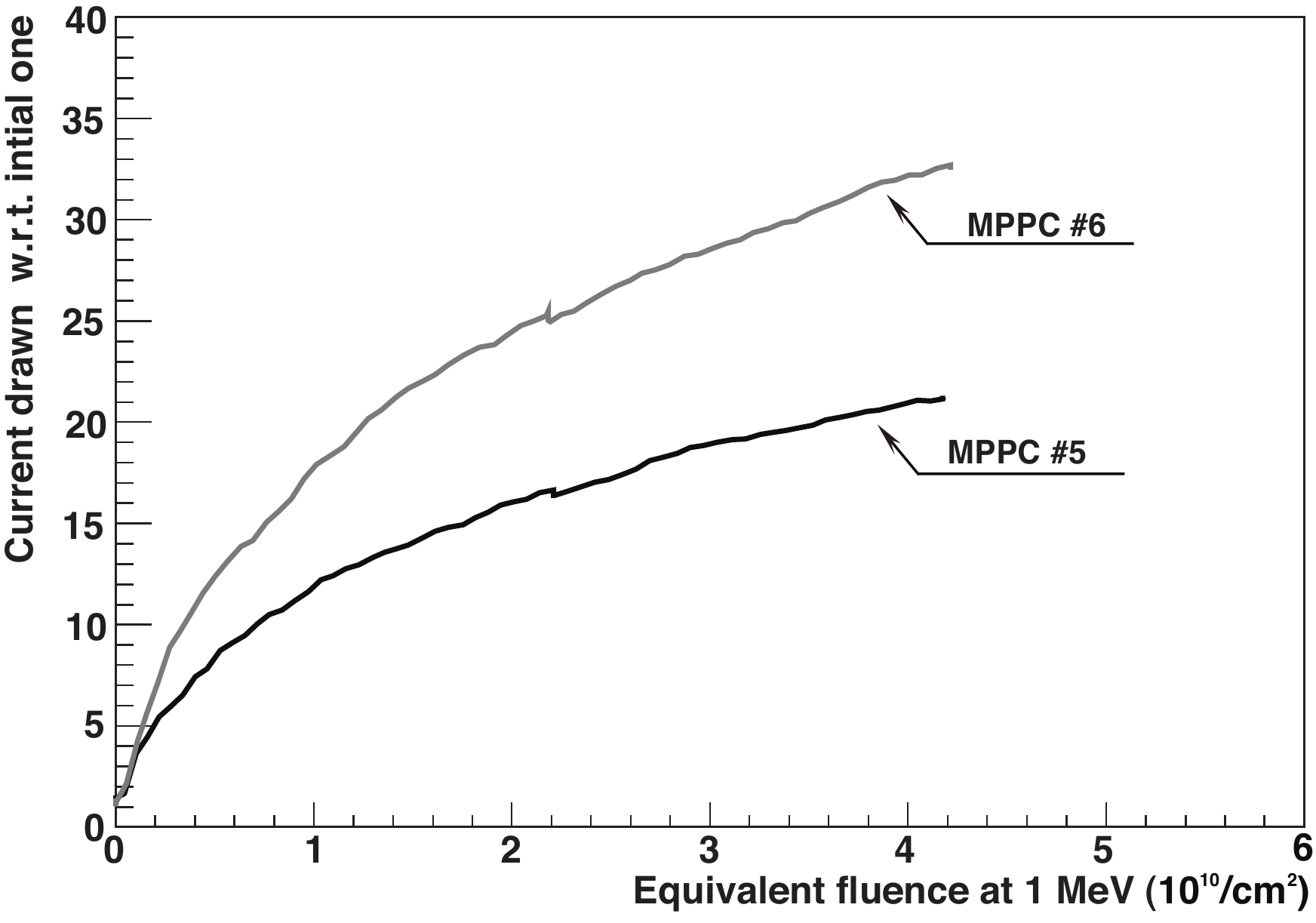, width=8cm}
\label{zoom}}
\caption{ Increase factors of the current drawn by the SiPMs (a) and the MPPCs (b) as a
function of the accumulated fluence.
\label{fig:curr_fluence}}
\end{centering}
\end{figure}

Data collected during the irradiation allow to investigate at which fluence the irradiation deteriorates significantly the detector performances. The measured dark currents drawn as a function of the irradiation time show significant differences among the devices depending on the neutron flux they are exposed to. Fig.~\ref{fig:curr_time} shows that during the four hours of data-taking some devices worsened by a factor 30, others by "only" a factor 10. These plots show that the effects which are synchronous among the devices, which could signal a sudden change in temperature, the effect of the night without irradiation, or similar, are small compared to the effects of interest.

It is then natural to look for a more general trend by considering the integrated fluence instead of the irradiation time. This is shown for the SiPMs in Fig.~\ref{fig:curr_fluence}a: the dark current increases monotonously ever since 10$^9\neq$ for all devices and independently of the neutron fluences that are different among them.  The curve of the relative increase is universal among SiPMs, including the larger area one: it can be fit with $f(x)=Ax^{2/3}$ where $A\sim 8.5$  and it shows that after a fluence of $10^{10}\neq$ ($4\times 10^{10}\neq$ )the current drawn is worsened by approximately a factor 10 (20). Similarly MPPCs (see Fig.~\ref{fig:curr_fluence}b) show a rapid degradation with irradiation, but the slopes of the dark currents can be significantly different.

\begin{figure}[hbt]
\begin{centering}
\subfigure[]{\psfig{file=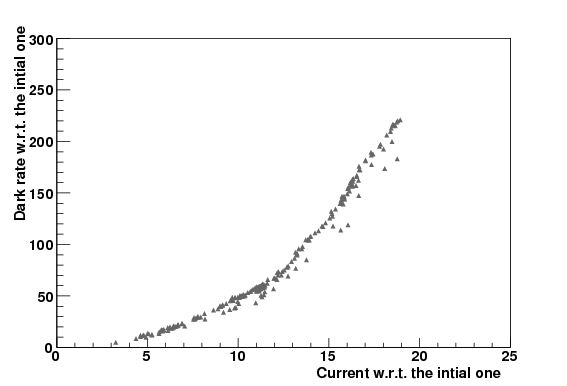,width=8cm}
\label{fig:ratevscurr}}
\caption{Dependence of the dark rate degradation factor on the current one for SiPM\#8 during the first irradiation run.}
\end{centering}
\end{figure}

Also, Fig.~\ref{fig:ratevscurr}) shows the dark rates obtained by setting a threshold at 2.5
photo-electrons
 as a function of the drawn current for a single run.  One can notice that the increase in rates is significantly larger than in currents. This can be an indication that the rate is mostly due to accidental coincidences between discharges of the individual pixels. While the drawn current is linear in the the rate of the individual discharges, the rate of accidental coincidences is not.

\begin{figure}[hbt]
\begin{centering}
\subfigure[]{\psfig{file= 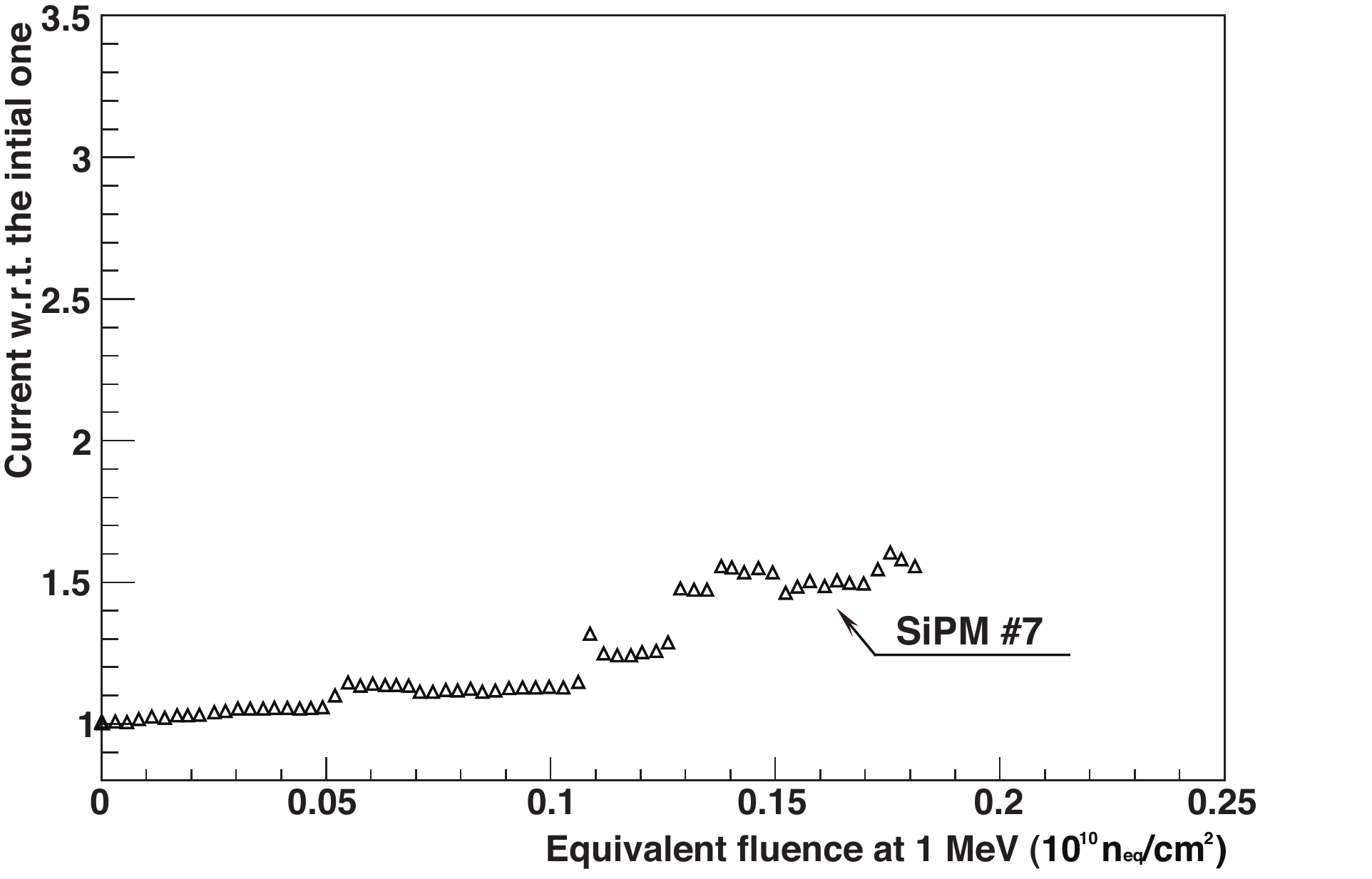,width=5cm}
\label{curr_low}}
\subfigure[]{\psfig{file= 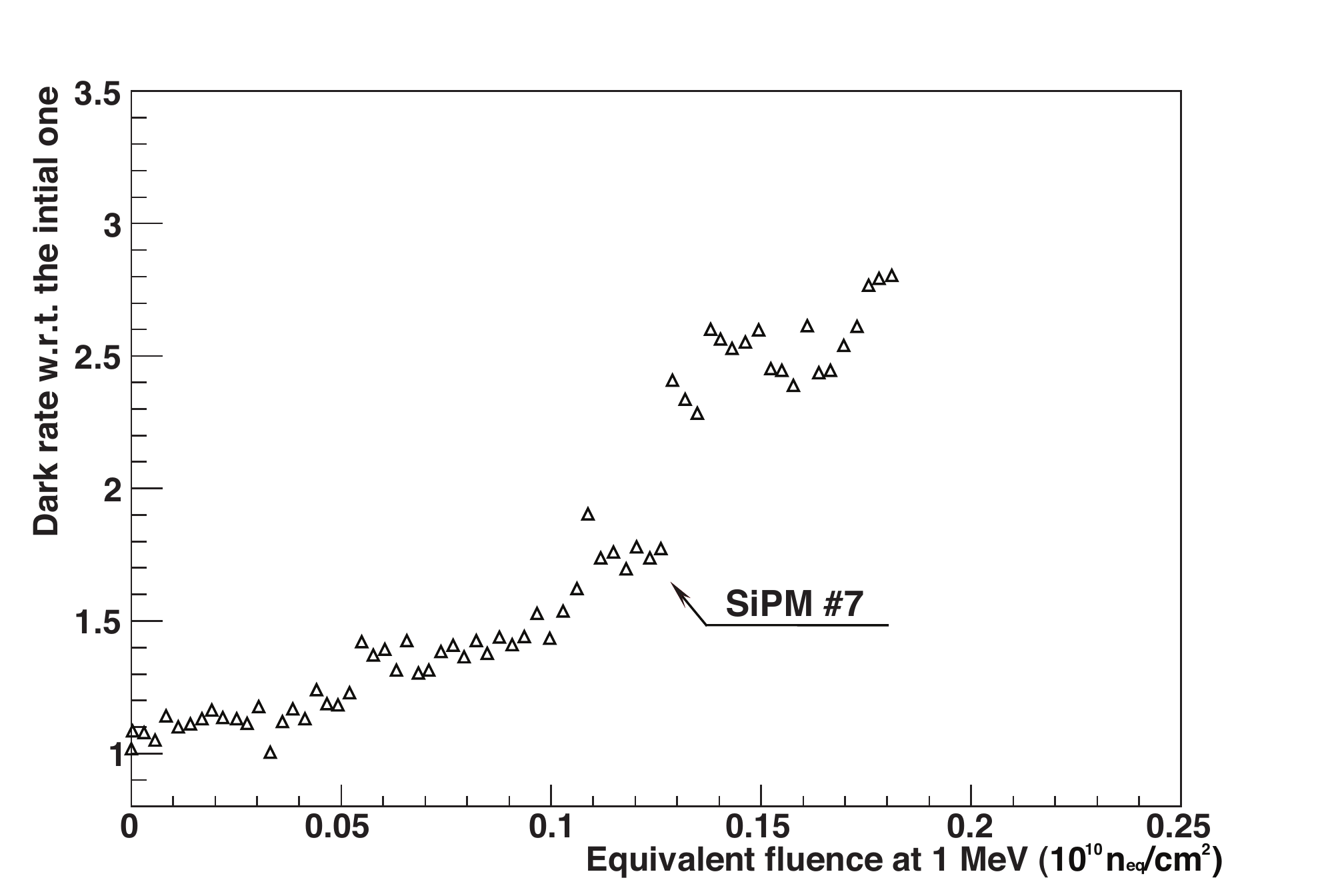, width=5cm}
\label{rh_low}}
\caption{ Increase factors of the current drawn by the SiPM as a function of the fluence (a) and of the corresponding dark rate for a threshold at approximately 2.5 photo-electrons  (b).}
\label{fig:curr_lowfluence}
\end{centering}
\end{figure}

To probe the resistance of a SiPM to lower fluences, one of the devices was also irradiated at a larger distance from the source, thus reducing the flux by about two order of magnitudes.  Fig.\ref{fig:curr_lowfluence}, shows that currents and dark rates are 
 stable up to fluences of the order of 2-4$\times$10$^8\neq$ and then,  they start to increase.
No significant recovery effects appeared after a whole night
without irradiation: the absolute value of 
the current and the increase rate, once the flux was back on, didn't change.  

\begin{figure}[t]
\begin{centering}
\psfig{file= 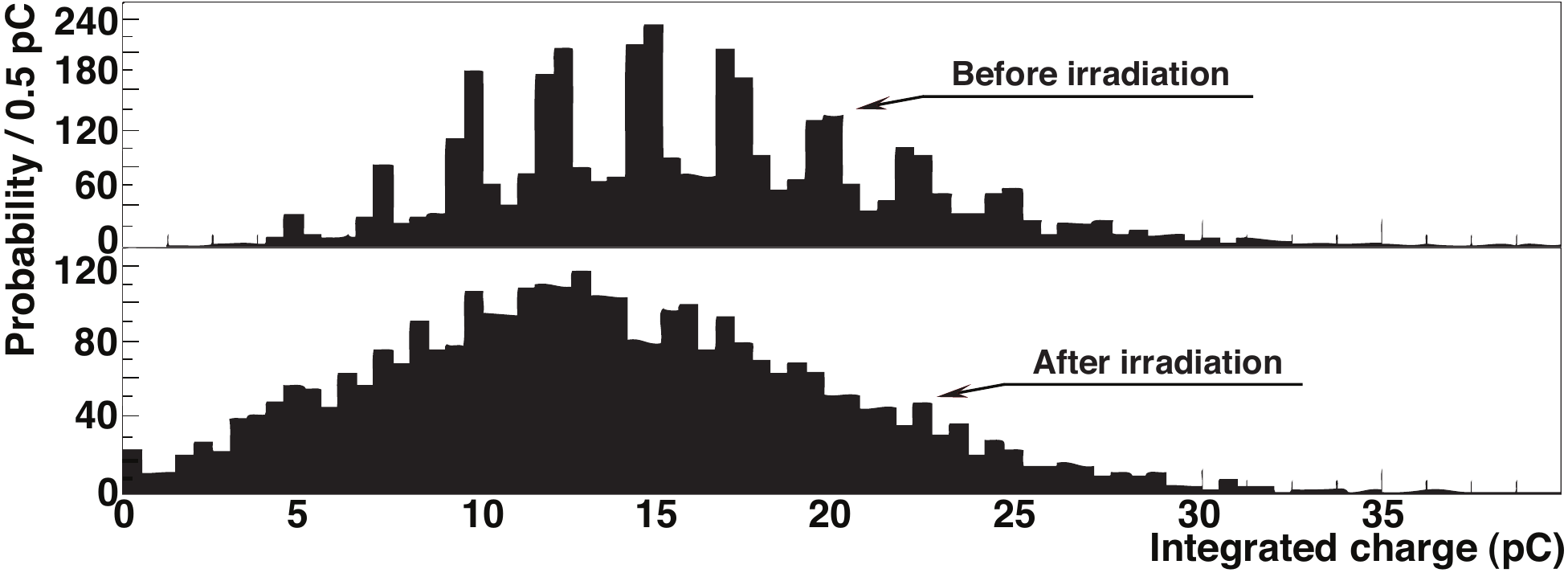,width=10cm}
\caption{MPPC charge spectra for a low intensity LED run, before (top) and after (bottom) irradiation. The irradiation light is arbitrary.}
\label{fig:singlephoton}
\end{centering}
\end{figure}

\section{Gains}
The effect of the irradiation on the gain was studied by testing the response of the Photo-Multipliers to a pulsed  LED yielding a low number of photo-electrons and to the  light produced by cosmic rays in a plastic scintillator, where the light yield is closer to the expected operation of the devices and reproducible.

The collected charge is measured by means of an ADC module with 0.25pC resolution, by using a gate window of about 30ns. The trigger is given by the pulse generator during the LED irradiation and by a pair of scintillators placed one above and one below the test module, during the cosmics run. 

\begin{figure}[hbt]
\begin{centering}
\subfigure[]{\psfig{file= 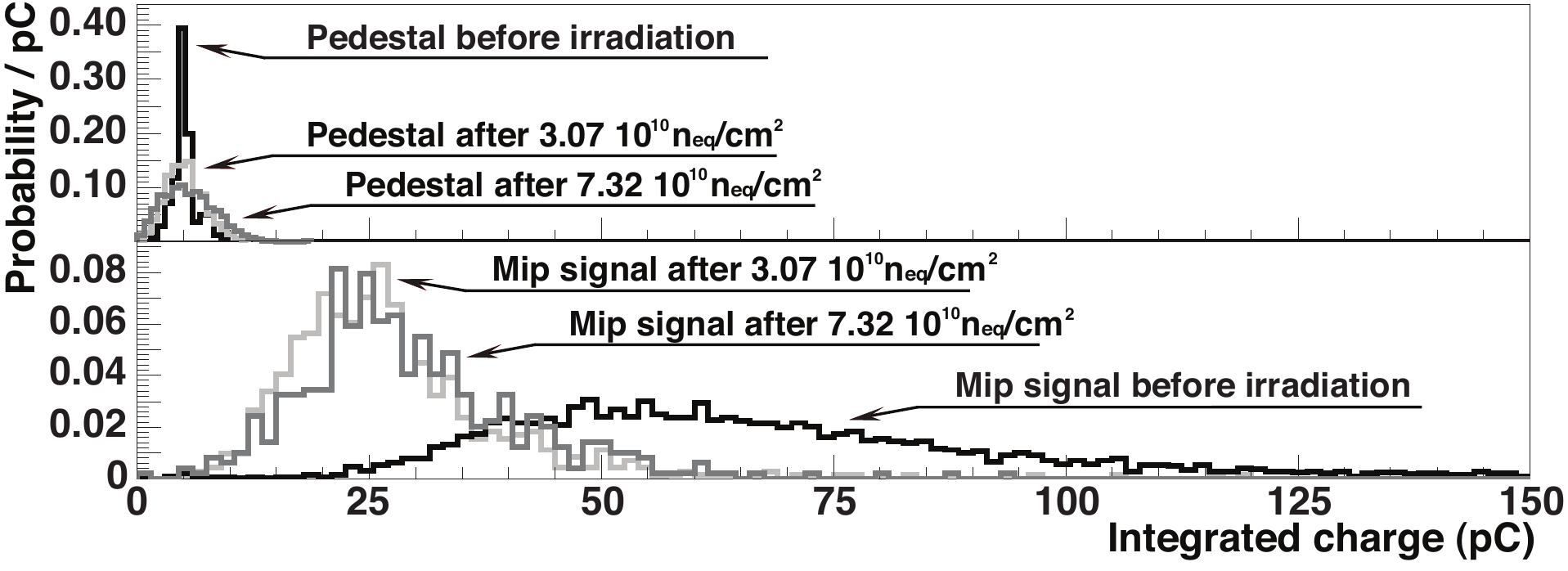,width=9cm}}
\subfigure[]{\psfig{file= 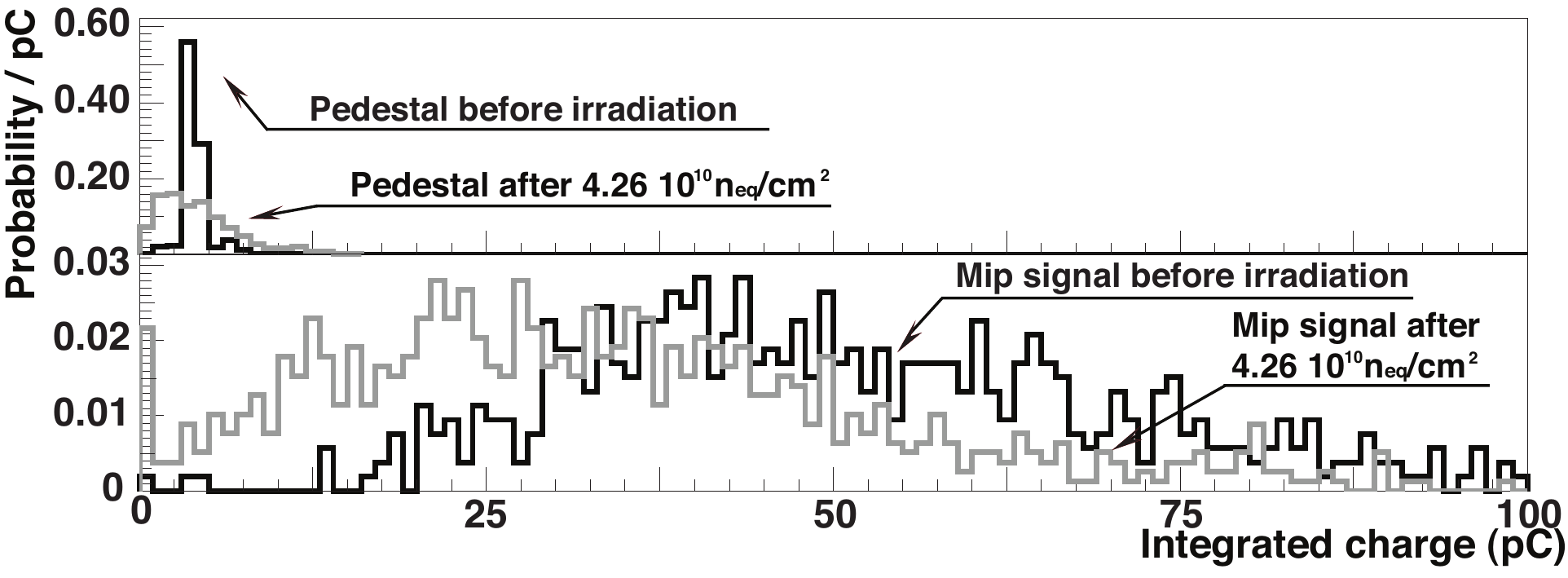,width=9cm}}
\caption{Pedestal (top) and cosmic signal charge spectra (bottom) for a SiPM (a) and an MPPC (b)
downstream of a 150 cm long WLS fiber for different integrated fluences.}
\label{fig:SiPM_cosmics}
\end{centering}
\vspace{-5pt}
\end{figure}

LED runs performed after irradiation show an almost complete degradation of the single-photon resolution due to the increase of the noise (see Fig.~\ref{fig:singlephoton}). Cosmic-ray runs allow instead to study the impact on the devices of the irradiation in terms of global gain (see Fig.\ref{fig:SiPM_cosmics}). Pedestals are stable in mean value, but broaden approximately by a factor three due to the increase of the dark rate intrinsic noise. Conversely the average gain of the irradiated devices lowers by approximately a factor two after irradiation, probably due to dead time increase on the individual pixels.

These effects lead in the cosmic-ray run to an important reduction of the detection efficiency of requiring a signal three standard deviations above the pedestal from more than 95\% to about 70\%.
No evident dependence of the performance deterioration on the integrated fluence was found.

\section{Conclusions}

Several Silicon Photo-Multipliers were exposed to an intense 
neutron flux integrating up to a total fluence of 7.32$\times$10$^{10}\neq$. 
Their performance were for the first time studied before, during  
and after the irradiation thanks to the use of a controlled neutron source (the ENEA FNG). The drawn currents were found to increase up to a factor 30 while the dark counts raise at a significantly worse rate. The detection efficiency, measured with cosmic rays, 
drops from above 95\% to around 75\%.

The measurements show that Silicon Photo-Multipliers performances deteriorate with irradiation even after  
few $10^8\neq$.
 A dedicated experiment at so low rates is being planned in order to better quantify the break-down fluence, if any.

\bibliographystyle{ws-procs9x6}

\end{document}